\documentclass[%
 aip,
 pof,
 sd,%
 amsmath,amssymb,
preprint,%
]{revtex4-1}

\usepackage{graphicx}
\usepackage{dcolumn}
\usepackage{bm}
\usepackage[euler]{textgreek}

\newcommand{\etal}{\textit{et al.}}
\newcommand{\ie}{\textit{i.e.}}

\begin{document}

\title{Spray formation: a numerical closeup}

\author{Yue Ling}
\affiliation{Institut Jean le Rond d'Alembert, Sorbonne Univ, UPMC Univ Paris 06, CNRS, UMR 7190, F-75005, Paris, France}

\author{Daniel Fuster}
\affiliation{Institut Jean le Rond d'Alembert, Sorbonne Univ, UPMC Univ Paris 06, CNRS, UMR 7190, F-75005, Paris, France}

\author{Gretar Tryggvason}
\affiliation{Department of Aerospace \& Mechanical Engineering, University of Notre Dame, Notre Dame, IN 46556, United States }

\author{St\'{e}phane Zaleski}
\affiliation{Institut Jean le Rond d'Alembert, Sorbonne Univ, UPMC Univ Paris 06, CNRS, UMR 7190, F-75005, Paris, France}
\email{stephane.zaleski@upmc.fr}

\begin{abstract}

  Spray formation and atomization in a gas-liquid mixing layer is an
  important fundamental problem of multiphase flows. It is highly
  desirable to visualize the detailed atomization process and to
  analyze the instabilities and mechanisms involved, and massive
  numerical simulations are required, in addition to experiment. 
  Rapid development of numerical methods and computer technology
  in the past decades now allows large-scale three-dimensional direct numerical
  simulations of atomization to be performed. Nevertheless, the
  fundamental question, whether all the physical scales involved in
  the primary breakup process are faithfully resolved, remains
  unclear. In the present study, we conduct direct numerical
  simulations of spray formation in a gas-liquid mixing layer with 
  state-of-the-art computational resources (using up to 4 billion
  cells and 16384 cores), in order to obtain a high-fidelity numerical
  closeup of the detailed mechanisms of spray formation. We also aim
  to examine whether present computational resources
  are sufficient for a fully resolved direct numerical simulation of
  atomization.

\end{abstract}

\maketitle

\section{introduction}
The breakup of fluid masses is a phenomenon of enormous complexity,
with diverse physical setups and mechanisms. When the
fluid masses break rapidly into large numbers of small droplets one speaks of
atomization \cite{Lasheras_2000a,Eggers_2008a}. Such atomization in
a gas-liquid mixing layer, where a high-speed
gas stream emerges from an orifice parallel to a lower-speed liquid
stream, has been studied in great detail \cite{Villermaux_2004a,Marmottant_2004a}. 
The resulting Kelvin-Helmholtz instability generates large coherent structures  
that grow in size as they propagate downstream, together with
equally growing wave-like structures \cite{Hoepffner_2011a} on the
liquid-gas interface.  The standard picture of atomization
\cite{Lasheras_2000a} is that two-dimensional wave structures form near
the orifice, develop into sheets, which in turn develop Taylor-Culick end rims.
The flow then becomes more markedly three  dimensional:
finger branching from the end rim, and then
various thread, fibers or ligament-like structures parallel to the flow
appear, which eventually break into droplets. This sequence
and its variants are called {\em primary} atomization, which
is supposed to be
followed by {\em secondary} atomization, 
the breakup of large drops further downstream whenever they 
interact with sufficiently high-velocity gas flow. 
Several types of probability distributions of droplet sizes have
been proposed and compared to experiments
\cite{Villermaux_2004a,Villermaux_2007a}. 
Another mechanism for primary
atomization is the formation of holes in the thin-sheet-like
structures that appear in the waves prior to the formation of
ligaments and fingers.  These holes-in-thin-sheets structures are quite
similar, but not identical, 
to the holes that form in bag-breakup secondary atomization \cite{Opfer_2014a}
and droplet impact \cite{Marston_2016a}.
{The hole formation has not been visualized as frequently in primary atomization
and is thus less firmly documented. }

In order to better understand the mechanisms underlying atomization, 
experimentalists have switched from
the coaxial round jets typical of industrial applications to 
a quasi-planar setup that is more favorable for detailed analysis 
\cite{Matas_2011a, Jerome_2013a, Matas_2015a}.
This setup has allowed precise measurements and detailed visualizations of the
droplet-forming process. In the quasi-planar configuration it is possible to
compare the growth and frequency of the Kelvin-Helmholtz instability in the 
linear regime as predicted by numerical simulation, linear stability theory
and experiments \cite{Fuster_2013a}.
Three-dimensional analysis is, for obvious reasons 
\cite{Gorokhovski_2008a,Lebas_2009a,Shinjo_2010a}, less advanced,
despite a large number of results in the references already cited. 
In this work we simulate a model of the quasi-planar experiment of 
Matas \etal \cite{Matas_2011a} in order to better understand the mechanisms of droplet formation. 

\section{Numerical method and simulation setup}
The computational setup is shown in Figure \ref{fig:general_behavior}. 
The domain is a box of dimensions $L_x \times L_y \times L_z$, 
where we inject two streams, liquid and gas, separated by a solid separator
plate of size 
$\ell_x \times \ell_y \times L_z$. The streams enter
through the boundary at $x=0$  
with velocities $U_l$ and $U_g$, respectively.  
The thickness of the liquid stream is $H$ while that for the gas stream is $H - \ell_y$.
The thickness of the boundary layers on the liquid and gas sides of the separator plate
are taken to be identical and denoted by $\delta$, and we take $\delta=H/4$.
The length $\ell_x$ and the thickness $\ell_y$ of the separator plate are 
$H/2$ and $H/32$, respectively and it has been shown that the details of the separator plate is 
immaterial to the atomization process as long as $\ell_y\ll \delta$. \cite{Fuster_2013a}
The domain is initially filled with gas and then liquid progressively enters it. 
In order to minimize 
the effect of the finite size of the domain, the dimensions of the box are large
in the $x$ and $y$ dimensions $L_x = 16 H$ and $L_y = 8 H$ (while $L_z$ is set to $2H$). 
Special care is taken to specify exit conditions to minimize the recirculating flow and avoid
excessive reinjection of coherent structures near the inlet. 

It is not possible with the present computational capability and numerical 
methodology to perform direct numerical simulations (DNS) in this setup using
the physical parameters (such as the large liquid-to-gas density ratio) exactly 
as in the experiments \cite{Matas_2011a}. This is due to the very wide range of relevant
physical scales. Indeed the
tiny submicron droplets generated are three orders of magnitudes smaller than $H$. 
To alleviate these problems we reduce the physical scale ($H=0.8$ mm is used here 
compared to $H=5$ to 20 mm in experiments \cite{Matas_2011a}) and 
choose a set of parameters that allows faster and easier simulations while still placing the flow
in the high-speed atomization regime. 
The physical parameters and the corresponding dimensionless parameters are given in 
Tables \ref{tab_phys} and \ref{tab}, using standard notations and international units.

\begin{table}[htbp]
  \centering
  \begin{tabular*}{1.0\textwidth}{@{\extracolsep{\fill} }cccccccccc} 
    \hline
    $\rho_l$ & $\rho_g$ & $\mu_l$ & $\mu_g$ & $\sigma$ & $U_{0,l}$ & $U_{0,g}$ & 
    $H$ & $\delta_g$	& $l_x$ \\
    ($kg/m{^3}$)& ($kg/m{^3}$) & ($Pa\,s$) & ($Pa\,s$) & ($N/m$) & ($m/s$) & ($m/s$)&
    ($m$) & ($m$) & ($m$) \\
    \hline
    1000 & 50 & $10^{-3}$ & $5\times10^{-5}$ & 0.05 & 10 & 0.5 & $8\times10^{-4}$ 
    & $2\times10^{-4}$ & $2.5\times10^{-5}$ \\
    \hline
  \end{tabular*}
  \caption{Physical parameters.}
  \label{tab_phys}
\end{table}

\begin{table}[htbp]
  \centering
  \begin{tabular*}{1.0\textwidth}{@{\extracolsep{\fill} }cccccc} 
    \hline
    $M$ & $r$         
& $m$  & $\mathrm{Re}_{g,\delta}$ & $\mathrm{We}_{g,\delta}$ & $\mathrm{Re}_g$\\
  $\rho_g U_g^2/(\rho_l U_l^2)$   &  $\rho_l/\rho_g$ 
& $\mu_l/\mu_g$  &  $\rho_g U_g \delta/\mu_g$  &   $\rho_g U_g^2 \delta/\sigma$ & $\rho_g U_g H/\mu_g$ \\
    \hline
  20 & 20 & 20 & 2000 & 20 & $8000$ \\
    \hline
  \end{tabular*}
  \caption{Key dimensionless parameters.}
  \label{tab}
\end{table}

We solve the Navier-Stokes equations for incompressible flow
with sharp interfaces and constant surface tension. 
The fields are discretized using
a fixed regular cubic grid (with cell size denoted by $\Delta$), and we use a projection method for 
the time stepping to incorporate the incompressibility condition. 
The interface is tracked using a Volume-of-Fluid (VOF) method with a Mixed Youngs-Centered Scheme 
to determine the normal vector
and a Lagrangian-Explicit scheme for the VOF advection  \cite{Tryggvason_2011a}. 
The advection of momentum near the interface is conducted in a manner consistent 
with the VOF advection \cite{Rudman_1998a} with the superbee limiter applied in flux calculation. 
The viscous term is treated explicitly. 
Curvature is computed using the height-function method \cite{Popinet_2009a}. 
Surface tension is computed from curvature 
by a well-balanced Continuous-Surface-Force method \cite{Francois_2006a, Popinet_2009a}. 
Density and viscosity are computed from the VOF fraction
by an arithmetic mean. To capture the dynamics of poorly resolved droplets accurately, 
droplets of size smaller than 4 cells are converted to Lagrangian point-particles \cite{Ling_2015a}. 
The overall method is implemented in the free code 
{\it PARIS} \cite{parissimulator} and validation studies can be found in Ling \etal \cite{Ling_2015a}.

To assess whether the present simulation is a full DNS of atomization, simulations are 
performed on four grids called M0, M1, M2, and M3 so that M$n$ has $H/\Delta= 32 \times 2^n$ 
points in the liquid layer $H$. 
For the M3 mesh, the simulation was performed on about 4 billion cells 
using 16,384 processors. The total simulation time for all four meshes took over 10 million CPU hours. 
The results presented correspond to the M3 mesh, unless stated otherwise. 

\begin{table}[htbp]
  \centering
  \begin{tabular*}{1.0\textwidth}{@{\extracolsep{\fill} } ccc ccc c @{}} 
    \hline
    Run & $\Delta$(\textmu m) & $H/\Delta$ &  Cells \#  & 
    Cores \#	& Total core-hrs\\
    \hline
    M0 &  25		&	32  	& $8.39\times10^6$	& 32		& $\sim 3\times10^3$\\
    M1 &  12.5  		&	64 	& $6.71\times10^7$ 	& 256		& $\sim 5\times10^4$\\
    M2 &  6.25   		&	128 	& $5.37\times10^8$  	& 2048		& $\sim 1\times10^6$\\
    M4 &  3.125   	&	256 	& $4.29\times10^9$ 	&16384	& $\sim 10\times10^6$\\
    \hline
  \end{tabular*}
  \caption{Summary of simulation runs.}
  \label{tab:runs}
\end{table}

\section{Results}
\subsection{Overall atomization process}
\begin{figure}[htbp]
\centering
\includegraphics[width=1.0\columnwidth]{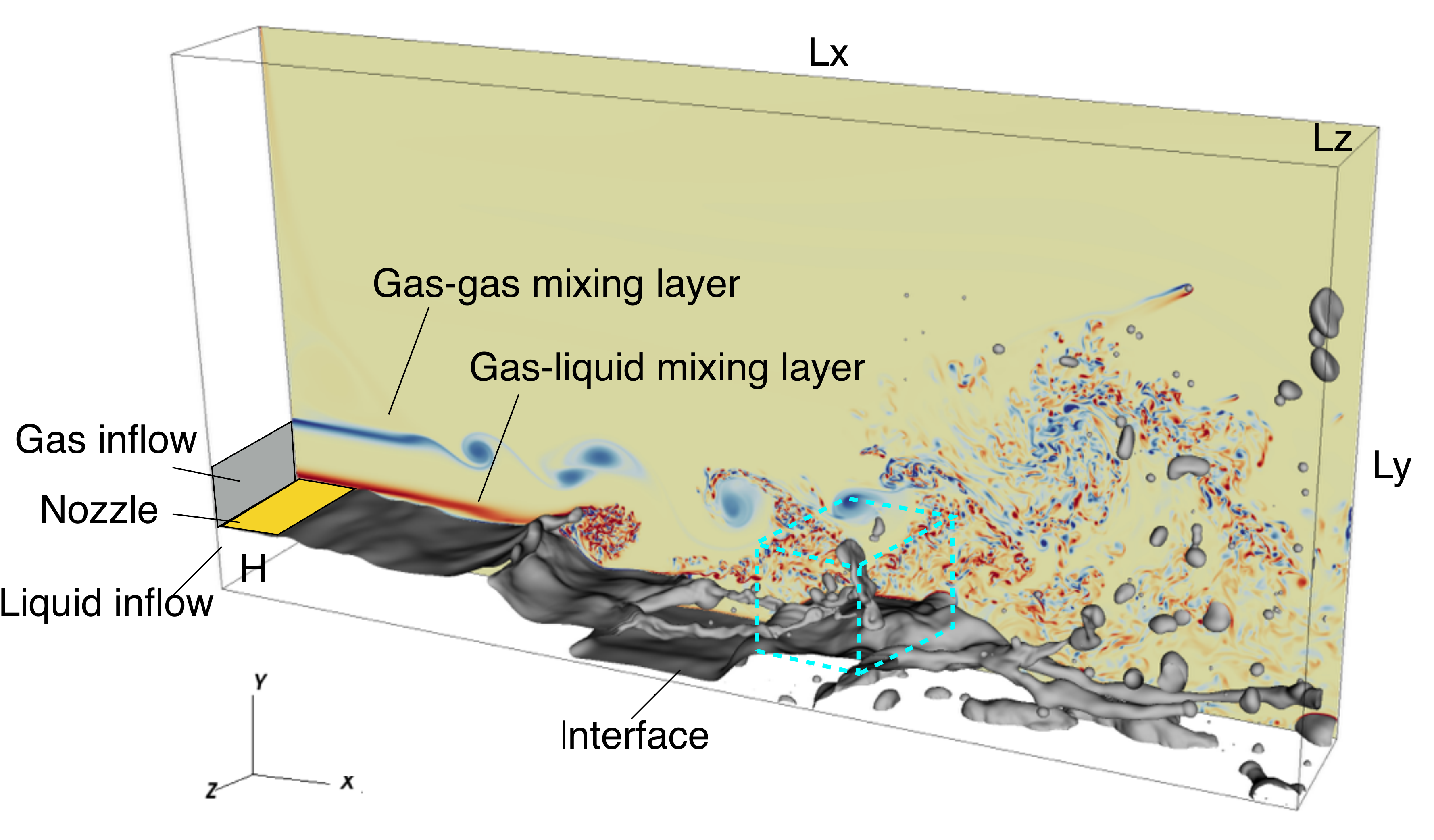}
\caption{Spray formation and atomization in a gas-liquid mixing layer. The $z$-vorticity is shown on the backplane. 
The sampling region for droplets statistics is indicated by cyan lines.
}
\label{fig:general_behavior} 
\end{figure}

A global view of the atomization in a gas-liquid mixing layer is shown in Fig.~\ref{fig:general_behavior}. 
The single-phase (gas-gas) and the two-phase (gas-liquid) mixing layers can be identified from 
the $z$-vorticity plotted on the backplane. Both of the mixing layers are unstable 
due to the velocity difference across the layers. 
The gas-liquid mixing layer develops faster and evolves a Kelvin-Helmholtz-like wave
on the interface. The interfacial wave grows and a thin liquid sheet forms at the wave crest. 
A Taylor-Culick rim appears at the edge of the liquid sheet. 
{The sheet folds and creases under the action of the gas turbulence, and this leads to 
perturbations of the rim. These perturbations 
produce small fingers which later develop into long ligaments. 
There is an important difference between the rim instabilities observed here 
and those seen for example in droplet splashes where interaction with energetic air motion is absent.}
 The ligaments eventually break into 
 small droplets due to Rayleigh-Plateau instability. The unbroken part of the liquid sheet 
reattaches to the domain bottom. 
Compared to the gas-liquid mixing layer, 
the gas-gas mixing layer evolves more slowly. The invasion of the turbulent vortices  
from the gas-liquid mixing layer accelerates the development the gas-gas mixing layer. 
Eventually, the two mixing layers merge and the downstream flows become fairly violent and chaotic. 

\subsection{Formation of sheets}
Due to the velocity difference between the gas and liquid across the interface, 
a Kelvin-Helmholtz-like wave develops on the interface and propagates downstream. 
As the interfacial wave grows the radius of curvature at the wave crest continues to decrease
and eventually liquid sheets form. 

\begin{figure}[tbp]
\centering
\includegraphics[width=1.0\columnwidth]{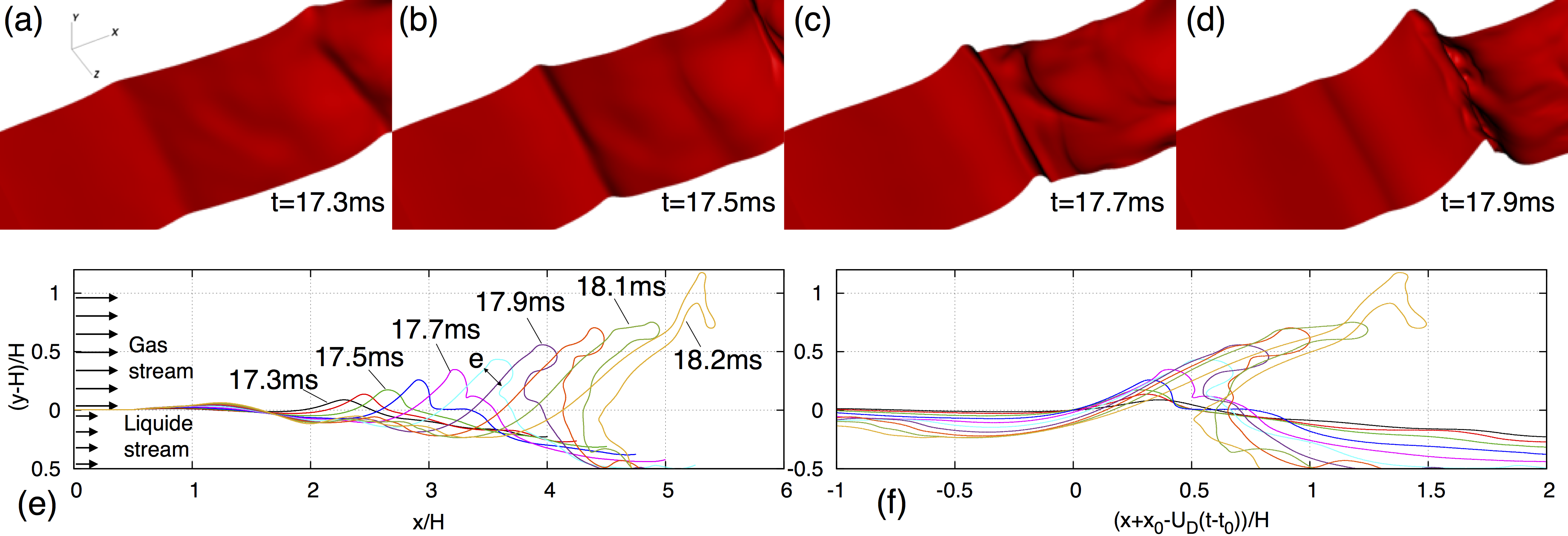}
\caption{Development of a quasi-2D interfacial wave, forming a liquid sheet: (a)-(d) time snapshots of the interface; interface profiles at plane $z=H$ with (e) original scale and (f) $x$-axis scaled by the Dimotakis speed $U_D$. }
\label{fig:2d-wave} 
\end{figure}

It is generally believed that the wave that appears first is a two-dimensional one
and then transverse instabilities (such as Rayleigh--Taylor (RT) and Rayleigh--Plateau (RP) instabilities) develop 
at the rim of the liquid sheet. This quasi-2D wave and its development are shown in Figs.\ \ref{fig:2d-wave} (a)-(d). 
The temporal evolution of the wave can be seen more clearly by a sequence of snapshots of the interface 
at the plane $z=H$ (see Fig.\ \ref{fig:2d-wave} (e)). 
The wave initially takes an Gaussian-like shape. The minimum radius of curvature is 
located near the wave crest, and decreases from 189.7 \textmu m at 17.3 ms to 43.1 \textmu m at 17.6 ms. 
Then the wave tends to fold forward. At a time between 17.7 and 17.9 ms, the two interfaces 
on both sides of the wave crest become parallel and form a liquid sheet. 
At this point, the thickness of the liquid sheet, denoted by $e$, is 174 \textmu m.  
As the sheet is pulled and stretched by the fast gas stream, its thickness decreases in time. 
{At $t=18.2$ ms (the last profile in Fig.\ \ref{fig:2d-wave}(e)) 
the minimum sheet thickness decreases to $e_{\min}=49.2\mu$m. At this scale the capillary
time is $\tau_{ca} = (\rho_l e^3/\sigma)^{1/2} \approx 0.05$ ms and the Ohnesorge number
is Oh$ = \mu_l (\sigma \rho_l e)^{-1/2} \approx 0.02$. There are still $e/\Delta = 16$ grid points for 
the sheet thickness. It is seen that at this time of $t=18.2$ ms the tip of the liquid sheet starts to fold.}
The radius of curvature at the hinge point is about 22 \textmu m. 
{The fact that the tip of the sheet folds, instead of forming a Taylor-Culick end rim
as expected for this time scale and Oh number, is a testimony of the strong interaction 
of the liquid sheet with the gas stream. 
The wave amplitude at $t=18.2$ ms is comparable to $H$ bringing the interaction to a maximum.}

The celerity of the interfacial wave is approximately a constant, which agrees well with 
the Dimotakis speed, 
\begin{equation}
	U_D=\frac{U_l+\sqrt{r}U_g}{1+\sqrt{r}}\, ,
\end{equation}
which is about 2.23 m/s for the present case. 
If the $x$-axis is shifted by $U_D$ with respect to the origin of wave formation $x_0$ and $t_0$, 
the waves at different times collapse, except the amplitude, as shown in Fig.\ \ref{fig:2d-wave}(f).
This is similar to the well documented 2D case \cite{Jerome_2013a}. 
 
\begin{figure}[tbp]
\centering
\includegraphics[width=1.0\columnwidth]{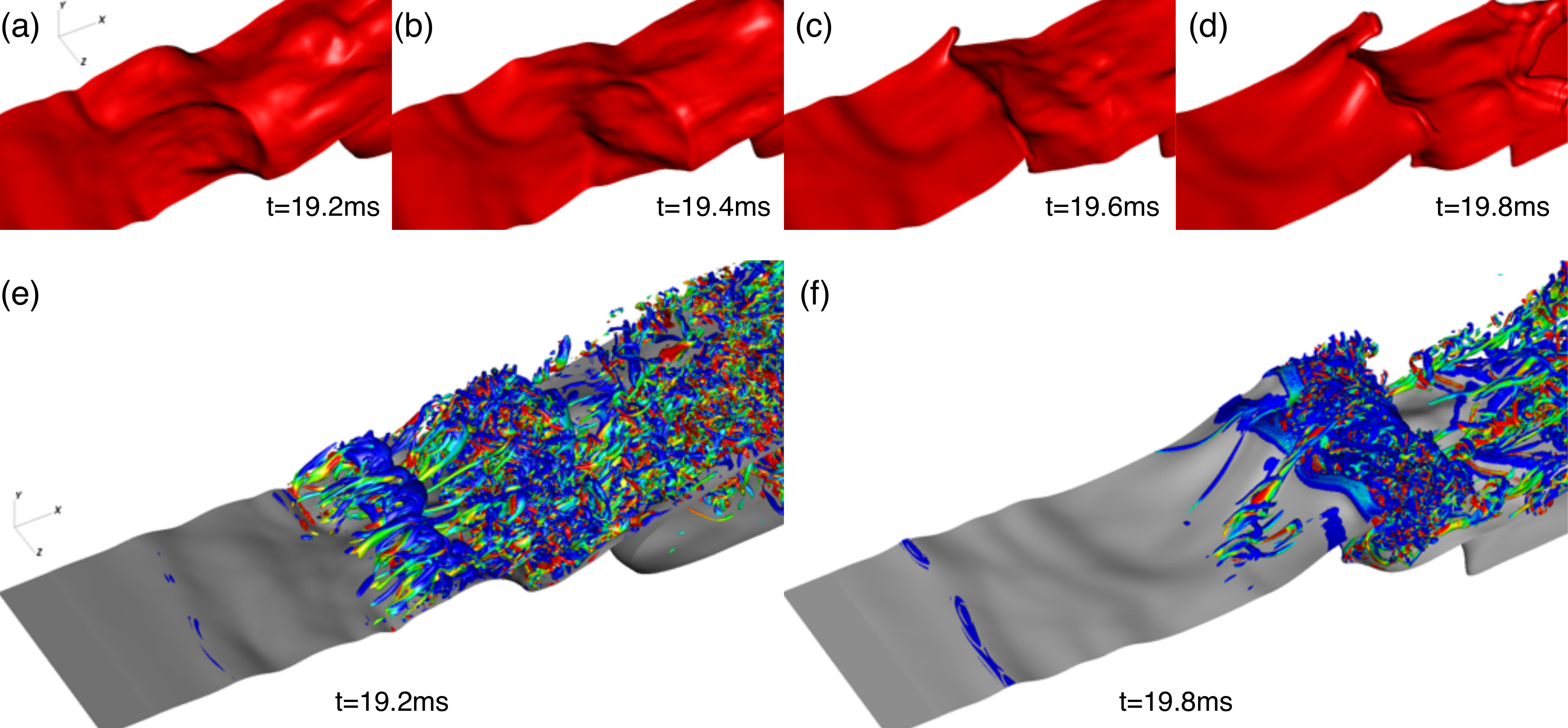}
\caption{Development of a fully 3D interfacial wave: (a)-(d) time snapshots of the interface and (e)-(f) turbulent vortical 
structures in the gas-liquid mixing layer (visualized by the $\lambda_2$ criterion). }
\label{fig:3d-wave} 
\end{figure}

Beyond the conventionally known quasi-2D waves, it is observed from the simulation results that 
the liquid sheet also forms in a fully three-dimensional manner (see Figs.\ \ref{fig:3d-wave} (a)-(d)), 
resulting in a significant transversely deforming rim at the sheet edge. This transverse wavelength is of the order 
of the width of the domain. (This may indicate that the domain width is too small
for the long wavelength modes in transverse instability.) 
It has been shown before that the Rayleigh-Plateau instability can induce transverse deformation 
of the rim, which later develops into fingers \cite{Roisman_2006a, Agbaglah_2013a}. 
%
%
%
%
%
%
%
%
%
%
{However, here the formation of 3D structures is clearly much faster than the Rayleigh-Plateau rim instability 
would be, and even occurs before the rim is formed at about $t=19.6$ ms in  Fig.\ \ref{fig:3d-wave} (c).}

\begin{figure}
\centering
\includegraphics[width=1.0\columnwidth]{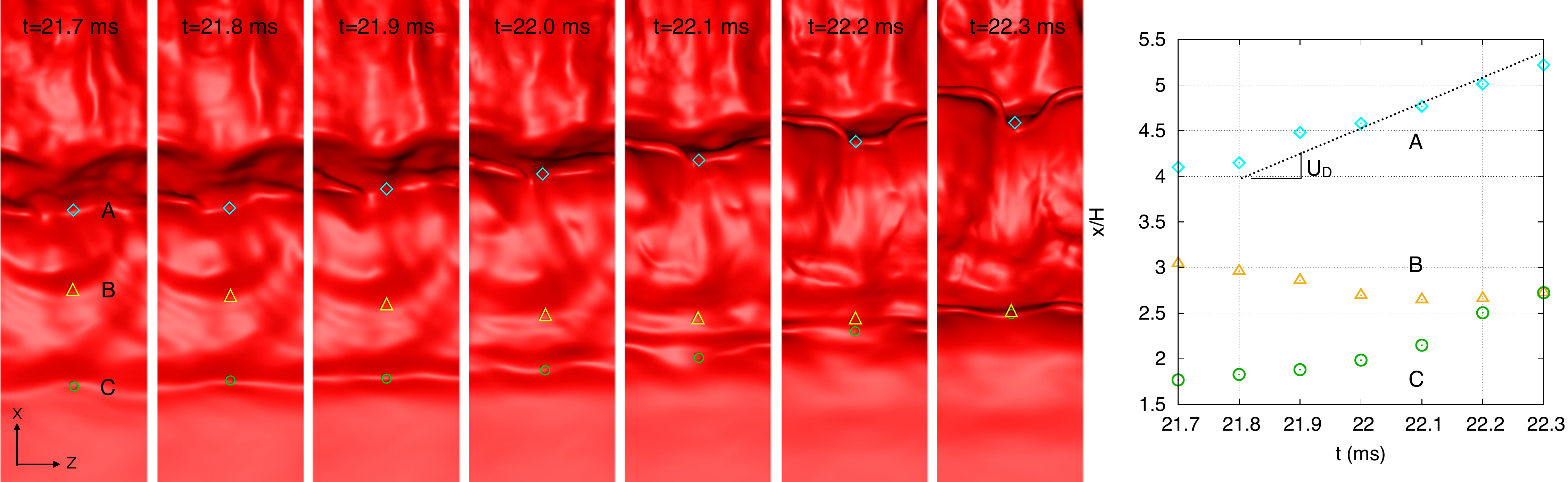}
\caption{Interfacial waves interaction (results by M2 mesh). Symbols indicate locations of the 
wave crests at plan $z=H$. The Dimotakis speed (black dashed line) is plotted for comparison.  }
\label{fig:interfacial-wave-interact} 
\end{figure}

Other different  mechanisms can contribute to the formation of the 3D wave. 
In particular, it has been shown by the transient growth theory that 
the 3D perturbations of a two-phase mixing layer can be more unstable 
than the 2D ones \cite{Yecko_2005a}. 
Furthermore, the turbulent gas flow on top of the interface impose significant 3D forcing
on the interfacial wave. See the turbulent vortical structures and their ``foot prints" on the 
interface in Figs.\ \ref{fig:3d-wave}(e) and (f). 
Finally, complex capillary wave interactions on the interface also contribute to triggering irregular 3D waves. 
As shown in Fig.\ \ref{fig:interfacial-wave-interact}, the capillary waves propagate 
both upstream (waves A and C) and downstream (wave B). The upstream and downstream propagating 
waves B and C meet and accelerate the development of the C wave. 

\subsection{Effect of mesh resolution}
\begin{figure}
\centering
\includegraphics[width=1.0\columnwidth]{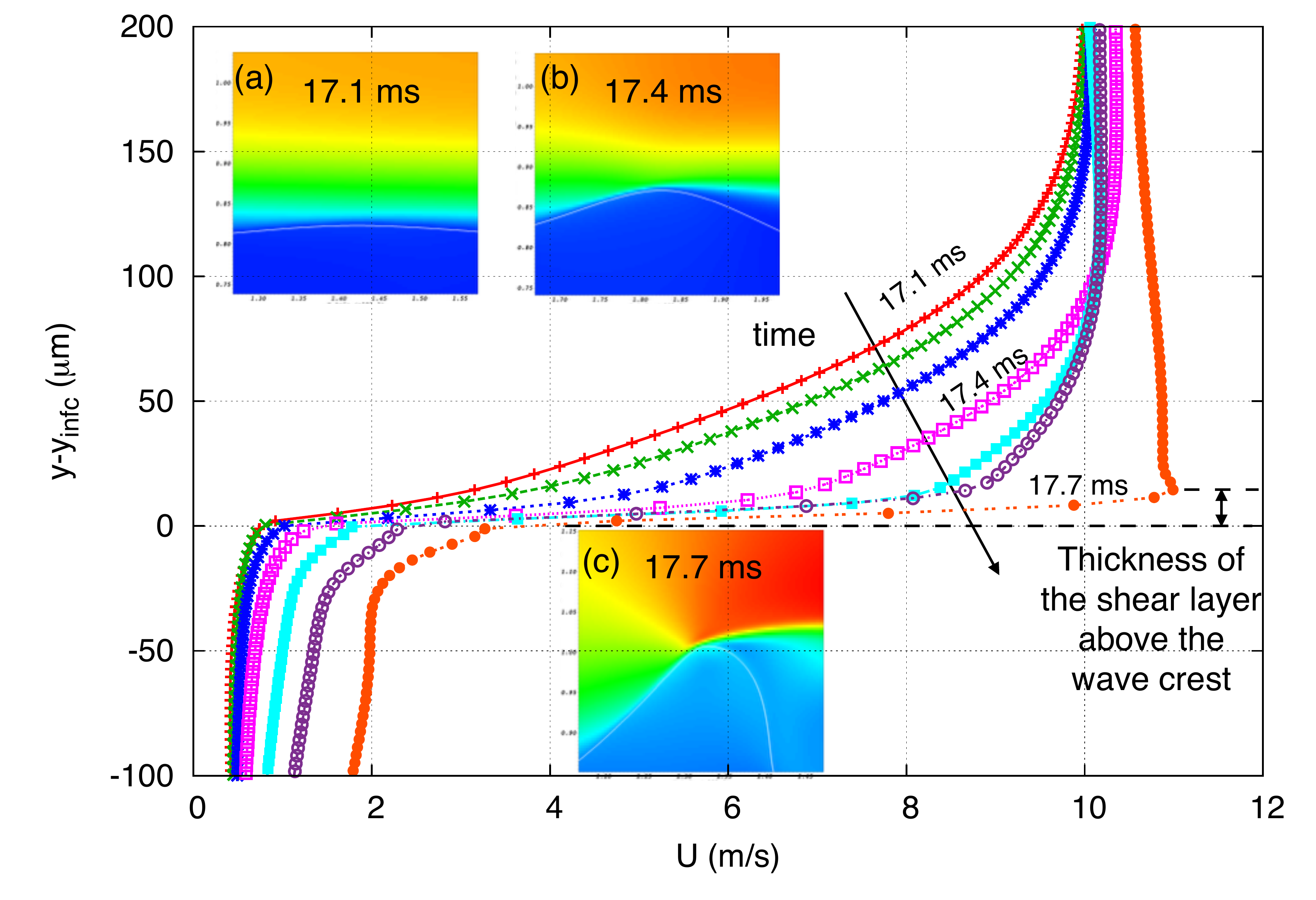}
\caption{Temporal evolution of the shear layer near the wave crest. 
(a)-(c) Streamwise velocity contours near the wave crest (gas-liquid interfaces are
indicated by white lines). 
The line plots are the streamwise velocity profiles in $y$ direction at the wave crest. 
The $y$ location of the interface is denoted by $y_\mathrm{infc}$. }
\label{fig:wave_crest} 
\end{figure}

\begin{figure}
\centering
\includegraphics[width=1.0\columnwidth]{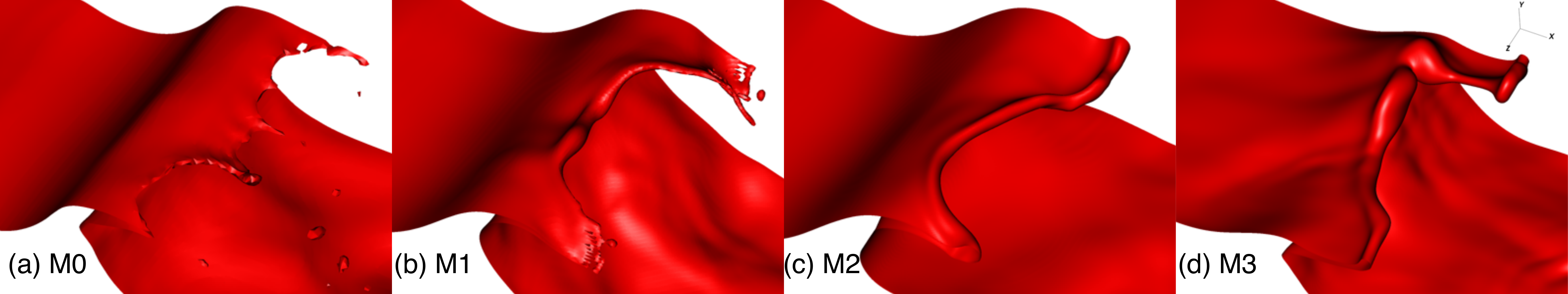}
\caption{A closeup at sheet formed at the wave crest for different mesh resolution.}
\label{fig:sheet_mesh} 
\end{figure}

It has been shown previously that the boundary layer of the injected gas stream must be 
well resolved, since otherwise the gas-assisted atomization and the frequency 
of the interfacial instability will not be accurately captured \cite{Fuster_2013a}. 
In the present study, we found that requiring sufficient numerical resolution to 
compute the formation of the sheet and the rim
indeed introduce a stricter  requirement on mesh size. 
As shown in Fig.\ \ref{fig:2d-wave}(b), the radii of the wave crest can go down to 
about 43.1 \textmu m ($t=17.6$ ms) or even lower to 22 \textmu m when the sheet folds,
%
%
which is much smaller than the injected gas boundary layer thickness $\delta$ ($\approx 200\mu$m). 
Furthermore, the thickness of the shear layer above the wave crest significantly decreases as the 
wave develops, see Fig.\ \ref{fig:wave_crest}. 
The shear layer thickness is initially similar to the boundary layer thickness of 
the injected gas stream $\delta$ (see $t=17.1$ ms), 
then it drops rapidly as the wave grows to about 15 \textmu m at $t=17.7$ ms. 
The M0 mesh ($\Delta = 25$ \textmu m) is clearly insufficient to resolve the wave crest curvature and 
the shear layer, as a result, the formation of the sheet is not properly captured. 
As shown in Fig.\ \ref{fig:sheet_mesh}(a), the rim is completely missed and the tip of liquid sheet breaks erroneously,
forming numerous tiny ligaments and droplets. 
The result for M1 ($\Delta = 12.5$ \textmu m) is better but two sides of the rim are still 
poorly resolved. For M2 and M3 meshes ($\Delta = 6.25$ and 3.125 \textmu m),  
about 4 and 7 cells are used to resolve the minimum radius of the wave
and about 3 and 6 cells for the shear layer above the wave crest. 
As a result, the sheet formation and the rim dynamics are well captured, 
see Figs.\ \ref{fig:sheet_mesh}(c)-(d). In such cases, 
no fingers or droplets are formed at this early stage. 

\subsection{Formation of ligaments}
\begin{figure}
\centering
\includegraphics[width=1.0\columnwidth]{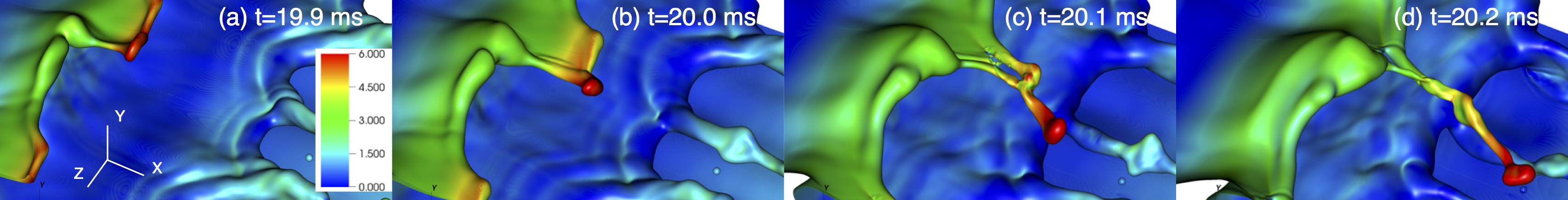}
\caption{Ligaments formation due to fingering at the tip of a liquid sheet. The color on the interface indicates the streamwise velocity. }
\label{fig:ligament_fingering} 
\end{figure}

The transverse instability of the rim is known to generate fingers at the tip of a liquid sheet \cite{Agbaglah_2013a}. The formation of a finger at the rim is well captured by the present simulation 
as shown in Fig.\ \ref{fig:ligament_fingering}.
The streamwise fluid velocity is also plotted on the interface and it is seen that the velocity increases 
gradually from the base to the round tip of the finger, 
indicating that the finger is stretched by the surrounding fast gas stream. Eventually the short finger develops into 
a long ligament, which breaks later to form droplets. 
%
%
%

\begin{figure}
\centering
\includegraphics[trim = 0 10mm 0 10mm, clip, width=1.0\columnwidth]{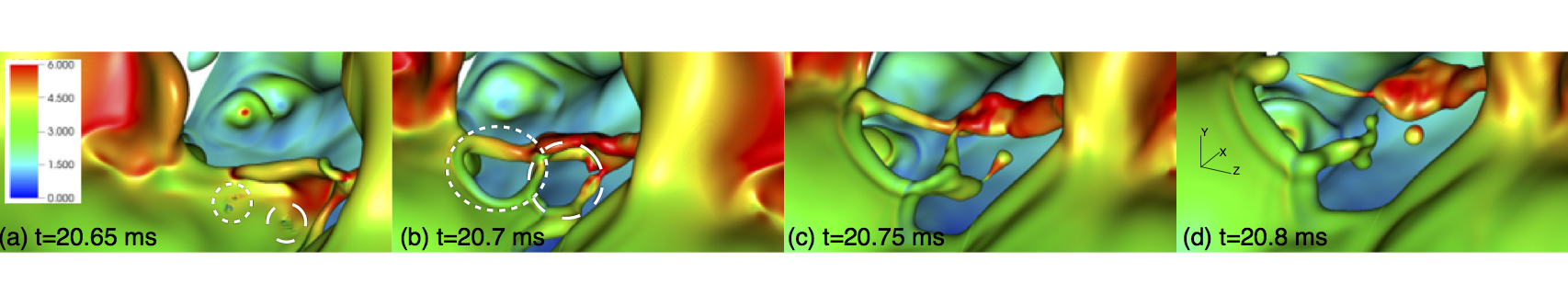}
\caption{Ligaments formation due to holes in a liquid sheet. The color on the interface indicates the streamwise velocity. }
\label{fig:ligament_hole} 
\end{figure}

Beyond fingering at the rim of the liquid sheet, holes appearing in the liquid sheet 
are observed to be another way to break the liquid sheet and to produce ligaments.
Similar to the fingers, the liquid sheet is also stretched by the gas stream and becomes 
thinner and thinner.  At a certain stage, holes are formed in the liquid sheet, see Fig.\ \ref{fig:ligament_hole}. 
The two holes are initially very small (highlighted by different dashed lines) 
but later they expand rapidly, causing the liquid sheet to rupture. 
Several small ligaments are generated and the orientations of these ligaments
are more diverse, different from the ligaments formed by fingering
which tend to align with the streamwise direction. 

For a stationary liquid sheet, holes are formed only when the sheet thickness is
very small ($e\sim O(10) \mathrm{nm}$) and the disjoining pressure becomes active. 
For a dynamic liquid sheet it has been shown in recent experiments that holes can form 
at a much larger thickness ($e\sim O(10)\mu$m) \cite{Opfer_2014a, Marston_2016a}.  
For example, in the experiment of Marston \etal \cite{Marston_2016a}, the sheet thickness
estimated by the hole expansion velocity and the Taylor-Culick theory is about 9 to 16 \textmu m. 
%
%
%
Several effects may explain the piercing of a liquid sheet at such large thicknesses, and among them
are Marangoni effects and perturbations from bubbles or droplets too small to be visible. 
%
%
Random perturbations from unseen objects are difficult to model, and Marangoni forces
are not included in the present simulation. Here
holes appear when the thickness of the liquid sheet decreases to about the cell size $\Delta$. 
This numerical cut-off length scale (the smallest $\Delta$ used is about 3.1 \textmu m) is 
much larger than breakup thickness of a stationary sheet {but is comparable or even smaller than 
the length at which dynamic liquid sheets are seen to break in experiments. }

\begin{figure}
\centering
\includegraphics[width=1.0\columnwidth]{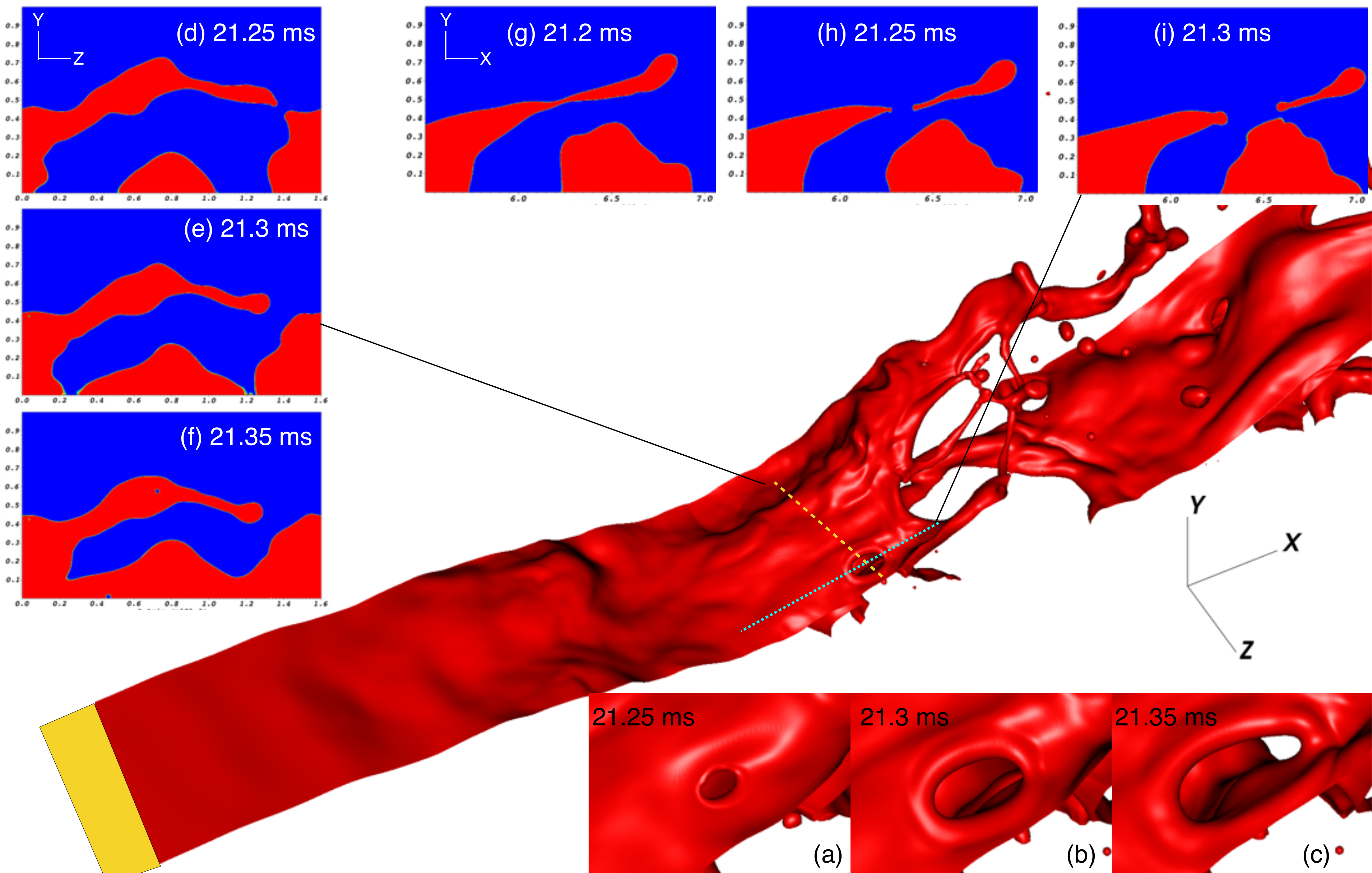}
\caption{Evolution of a hole formed in a liquid sheet. (a)-(c) A closeup a the hole expansion.
Liquid volume fraction (red) at $y-z$  and $y-x$ planes cutting through the holes are also shown
in (d)-(f) and (g)-(i), respectively.}
\label{fig:hole_expansion} 
\end{figure}

The evolution of a hole formed in a liquid sheet is shown in Fig.\ \ref{fig:hole_expansion}.
The measured hole expansion velocities in the streamwise and transverse directions 
are $U_{h,x}\approx 2.70$ m/s and $U_{h,z}\approx0.978$ m/s, respectively. 
As can be seen from the cross sections of the hole in the $y-z$ and $x-y$ planes, the sheet 
thickness near the hole is very uneven. The minimum sheet thickness just before the hole
appears is about 22 \textmu m, and after the sheet rupture the thickness in the vicinity of the hole varies
from 27 to 85 \textmu m. The Taylor-Culick velocity, 
\begin{equation}
	U_{h,TC} = \sqrt{\frac{2\sigma}{\rho_l e}}
\end{equation}
can be calculated based on the sheet thickness, yielding $U_{h,TC}=0.95\sim1.68$ m/s. 
It is seen that $U_{h,z}$ agrees well with $U_{h,TC}$. The excess of $U_{h,x}$ over the Taylor-Culick prediction 
is due to the streamwise stretching the liquid sheet, which causes the hole to expand faster in the $x$ than 
in the $z$ direction. 

Since mechanisms of sheet rupture, such as disjoining pressure, are absent in the present study, 
the initial formation of the holes is mesh dependent.
However, it is quite clear that the subsequent hole development and the rim around the hole are well resolved
with the M3 mesh. As a consequence, further increase of mesh resolution will only delay the pinch-off point 
but will not affect the ligaments formed from the expansion of the holes. 

\subsection{Formation of droplets}
\begin{figure}
\centering
\includegraphics[width=0.8\columnwidth]{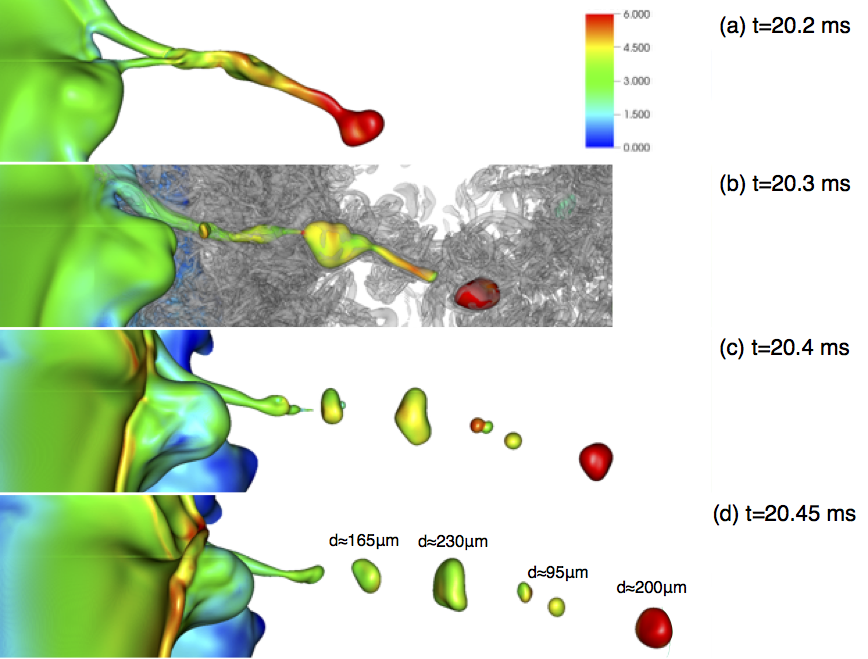}
\caption{Droplets generation due to ligament breakup. The turbulent vortices surrounding the ligament
is plotted in (b) by the $\lambda_2$ criterion.}
\label{fig:RPI} 
\end{figure}

Eventually the ligaments break into small droplets and one realization of the breakup process
is shown in Fig.\ \ref{fig:RPI}. The ligament here is the same one as shown in Fig.\ \ref{fig:ligament_fingering}. 
The ligament exhibits a very irregular shape compared to typical Rayleigh-Plateau 
breakup of a stationary ligament. The ligament diameter varies from 72 to 244 \textmu m along its axis.
The stretching by the surrounding turbulent gas stream (see Fig.\ \ref{fig:RPI}(b)) 
clearly contributes to the irregular breakup. 
The neck behind the tail of the ligament pinches off, forming a big droplet of $d\approx 200 \mu$m. 
The retraction of the ligament tail from the pinch-off point forms a big liquid blob in the middle of 
the ligament. Similar behavior is also observed in experiments \cite{Villermaux_2004a}. 
Coalescence of small droplets are also seen in Fig.\ \ref{fig:RPI}(c). 
At the end, a series of droplets varying from 95 to 230 \textmu m are produced. 

\begin{figure}[htbp]
\centering
\includegraphics[width=1.0\columnwidth]{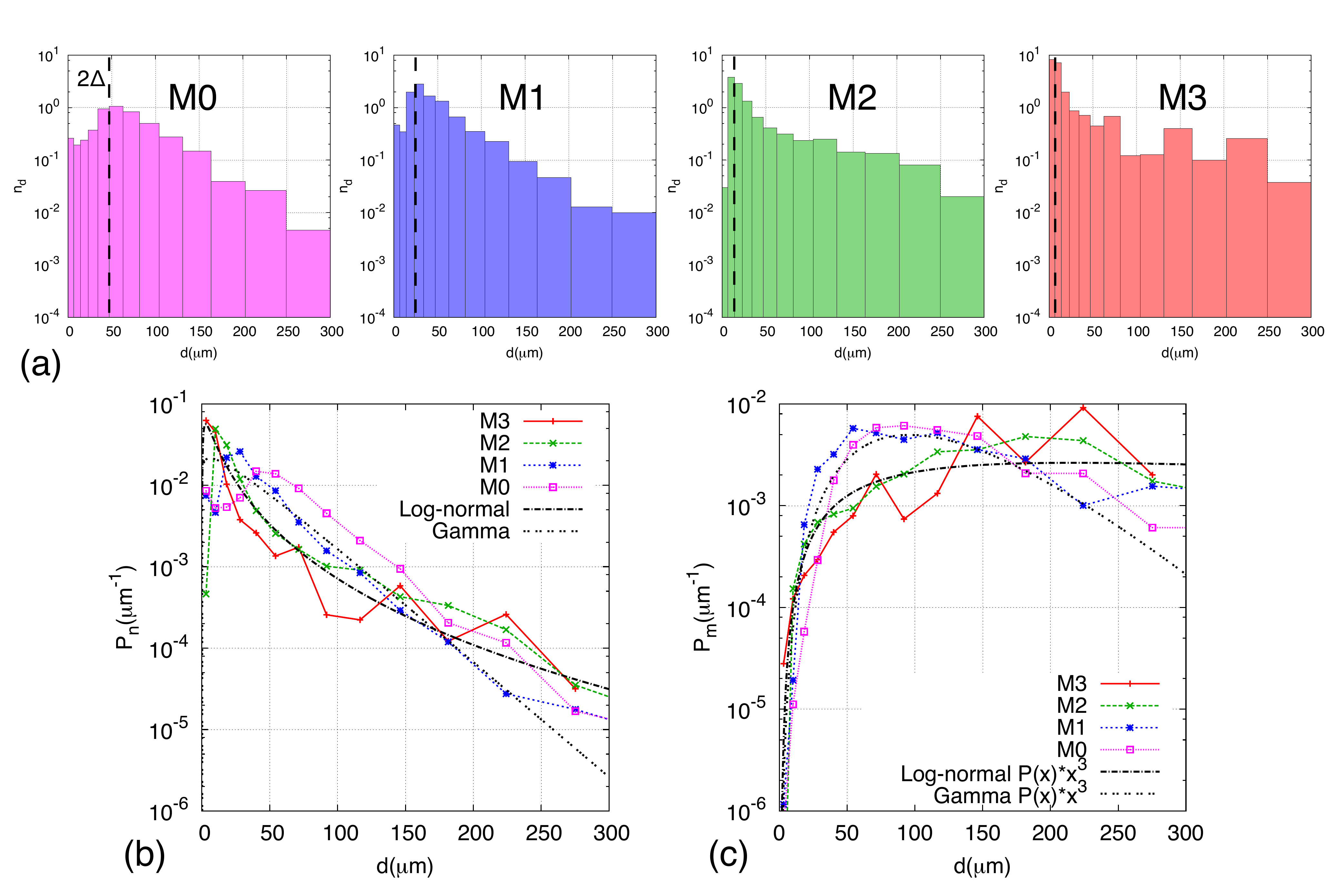}
\caption{The droplet size PDF obtained from different meshes (a) M3
(b) M2 and (c) M1 and (d) M0. The bin width for all the cases is 6.25 \textmu m. 
(e) Comparison of PDF profiles with Lognormal and Gamma distribution functions.   }
\label{fig:pdf} 
\end{figure}

To have a more general analysis of the droplets formed in atomization, we 
investigate the size distribution of droplets in a cubic box located downstream 
of the breaking wave. The sampling region is indicated in Fig.\ \ref{fig:general_behavior}.
The edge length of the cubic box is $2H$ ($8\le x/H \le 10$, $0.5 \le y/H \le 2.5$, and spanning
the whole width of the domain). 
The sampling is conducted after the atomization has reached a statistically steady state and 
over time intervals of 48, 21, 27, and 4.6 ms for the M0, M1, M2, and M3 meshes, respectively. 
(The sampling time for M3 mesh is significant shorter due to the extremely high computational cost.)
The average number of droplets $n_d(d)$ as a function of droplet volume-based diameter
is plotted in Fig.\ \ref{fig:pdf}(a), where $n_d(d)$ is defined as 
\begin{equation}
	n_d(d) = \frac{N(d)}{N_s}\,,
\end{equation}
where $N(d)$ is the total number of droplets collected within the bin centered at $d$ 
and $N_s$ is the number of samples.
The bin width is varied in $d$, starting from 6.25 \textmu m and then increasing by a constant ratio 1.2. 
The reason for using wider bins for larger $d$ is to reduce the fluctuations due to low numbers of 
larger droplets. Furthermore, it should be noted that the generation of droplets of size smaller than 2$\Delta$ 
for each mesh is quite likely not well captured in the present simulation. As a result, the droplets on the left of the dashed line ($d=2\Delta$) 
are less trustworthy. 

When the mesh size decreases, not only are more smaller droplets ($d\lesssim50\mu$m) 
captured (which is as expected), but we also observe that more larger droplets ($d\gtrsim100\mu$m) are collected. 
These large droplets ($d\gtrsim 100\ \mu$m), such as those shown in Fig.\ \ref{fig:RPI}, are typically generated 
from thicker ligaments, which are in turn produced by fingering at the end rim of the sheet (see Fig.\ \ref{fig:ligament_fingering}) or holes-induced sheet rupture (see Fig.\ \ref{fig:ligament_hole}). 
If the mesh is not sufficiently fine to capture the Taylor-Culick rim at the edge of a liquid sheet 
(or at the edge of a hole) as shown in Fig.\ \ref{fig:sheet_mesh}, 
then such a thick ligament may not get a chance to form. Instead, many tiny ligament will be produced
due to numerical breakup. 
As a consequence, less large droplets appear in the M0 and M1 results. 

The probability distribution function (PDF) of droplet number ($P_n$) and mass ($P_m$)
are shown in Figs.\ \ref{fig:pdf}(b) and (c), respectively, where  $P_n$ and $P_m$ are defined as 
\begin{equation}
	P_n(d) = \frac{N(d)}{\Delta_d \sum N(d)}\, 
\end{equation}
and 
\begin{equation}
	P_m(d) = \frac{m(d)}{\Delta_d \sum m(d)}\, 
\end{equation}
where $\Delta_d$ is the bin width and $m(d)$ represent the total mass of droplets collected in 
the bin centered at $d$.  
Due to the fact that more smaller ($d \lesssim 50\mu$m) and larger droplets ($d\gtrsim 100 \mu$m) 
are captured in fine mesh runs, $P_n$ is more convex for the fine meshes than for the coarse meshes.  
In spite of the discrepancy for the smaller and larger droplets,  
the slopes of $P_n$ (in a log-linear scale) for the four different meshes 
agree quite well for an intermediate range of droplet size between 50 to 100 \textmu m. 
In general, $P_n$ for M2 and M3 meshes agree reasonably well with each other, but  
are significantly different from the M0 and M1 results. The M3 PDFs are more noisy due to the 
shorter sampling time. To get a smoother PDF, 
the simulation has to be conducted for a longer time (for example, the M0 size distribution
is much smoother than the others due to the longer sampling time). 
However, due to the extreme computational cost for the M3 mesh, this task 
can only be relegated to future works. 

It is quite clear that the number of small droplets ($d\lesssim 50\mu$m) increases 
when the mesh is refined and thus is still mesh dependent, even for the finest mesh used here. 
However, these small droplets consist of only a small fraction of the total mass as shown in Fig.\ \ref{fig:pdf} (c). 
Compared to the droplet number distribution, for some applications the mass distribution
is indeed more important in characterizing sprays. It is observed that the 
main contribution to the mass is from large droplets ($d \gtrsim 50 \mu$m), 
which are well captured by the fine meshes like M2 and M3 used in the present study. 
The difference between the results of the fine and coarse meshes 
are more profound in $P_m$. The peaks for M0 and M1 are located at about 50-90 \textmu m; 
while those for M2 and M3 at about 180-230 \textmu m. The shift toward the larger droplets is due to 
the fact  that more larger droplets are captured by the finer meshes. 

Finally, the Log-Normal and Gamma distribution functions are employed to fit 
the PDF. The Log-Normal distribution functions is given as
\begin{equation}
	P_{L}({d}) = \frac{1}{{d} \eta \sqrt{2\pi}} \exp{\left[ -\frac{(\ln {d}- \zeta)^2}{{2}\eta^2} \right]}\, ,
\end{equation}
where the mean and the variance of $\ln{d}$ are $\zeta$ and $\eta^2$. 
The Gamma distribution can be expressed as 
\begin{equation}
	P_{G}({d}) = \frac{\beta^{\alpha}}{\Gamma(\alpha)} {d}^{\alpha-1} \exp \left(-\beta{d} \right)\, , 
\end{equation}
where $\alpha=(\hat{\zeta}/\hat{\eta})^2$ and $\beta=\alpha/\hat{\zeta}$,  
and the mean and variance of $d$ are denoted as $\zeta'$ and $\eta'^2$ . 

A comparison between the present simulation results and the PDF models 
are shown in Fig.\ \ref{fig:pdf}(b) and (c). 
The Log-Normal distribution is fit with $\eta \simeq 1.2$ and $\zeta \simeq 2.5$; while the
Gamma with $\alpha \simeq 1.2$ and $\beta \simeq 0.033$.
The Log-Normal distribution is observed to match better the finer mesh results; 
while the Gamma distribution seems to fit better the coarser mesh results. 
The Log-Normal distribution has also been observed to well fit the droplet size distribution 
in recent experiments \cite{Marty_2015a}. Furthermore, it is interesting to notice that $\eta$ 
measured experimentally varies from 1 to 1.8, which agrees well with that obtained in the present simulation. 

In spite of the reasonable performance of the distribution models, the physical 
reasons behind are still not fully understood. 
The idea behind the Log-Normal model is that the formation of droplets is a sequential cascade of
breakups, where the larger mother drops break into smaller daughter droplets. The ratio between 
the daughter and mother drops in each breakup is a random fractional number which follows a normal distribution. 
As a result, the size of the droplets formed at the end follows a Log-Normal distribution. 
In contrast to the breakup process, the coalescence between droplets introduces an inverse 
cascade, \ie, smaller droplets collide and merge to form bigger droplets 
or the coalescence of the smaller blobs constitutive of a ligament forms bigger blobs.
These aggregation scenarii will result in a Gamma distribution for the droplet size \cite{Villermaux_2007a}. 
The simulations presented here show that the spray formation is through a sequence of 
different complex mechanisms, and neither of these two PDF models 
are therefore likely to completely capture these mechanisms. 
We have observed breakups in sequence, the bulk liquid first breaks into thin liquid sheets , 
then the sheets into fingers and ligaments, and at last the ligaments into droplets of different size. 
However, the process of spray formation is clearly not a sequence of random breakups 
like suggested in the Log-Normal model. 
(We rarely see a droplet, once formed, further break into smaller droplets in the fine mesh runs.)
On the other hand, coalescences of droplets, the assumption behind the Gamma distribution model, 
are only occasionally observed. Therefore, it is also quite likely that the aggregation would not
have a significant impact on the droplet size, either. 

Notice that both of these classical distributions, the Log-Normal and the Gamma, are obtained
when a scaling process is observed, that is when nonlinear effects occur over a large range of scales.
For example, Kolmogorov turbulent cascade for the Log-Normal distribution, 
or Einstein-Smoluchowski aggregation dynamics for the Gamma distribution, 
both span over a wide range of scales from $\ell_{\min}$ to $\ell_{\max}$. The fact that $\ln(\ell_{\max}/\ell_{\min})$
is large is a condition of applicability of the central limit theorem in these theories. 
Here the best fit to the Log-Normal indicates that $\ln(\ell_{\max}/\ell_{\min}) \sim 2 \eta \approx 2.4$. 
Compared to many scaling observations performed in physics over a moderate range of scales, 
to be specific with just one decade as $\ln(\ell_{\max}/\ell_{\min}) \ge \ln 10 \approx 2.3$, 
the present system has a sufficiently large range of scales to consider scaling hypotheses, 
but not yet a ``truly" large range of scales as in Kolmogorov turbulent cascade experiments at
the large Reynolds numbers.
The absence of a truly large range of scales makes it difficult to draw definite conclusions from 
the fit of the droplet size probability distribution to the classical theories, 
but it also indicates that none of these theories is without doubt in its range of validity.

Another analysis of the droplet size distributions, that does not involve a single scaling process, is to consider
several distinct processes at different scales, for instance "rim drops" from the disintegration of the Taylor-Culick rims
and "film drops" from the disintegration of the thin sheets. In some experiments, \cite{Marmottant_2004a}  hints of the bimodal
distributions that would result from two distinct processes have been seen. 
We observe no such effects in our distributions,
and believe that the oscillations of the M3 distribution are entirely explainable by the statistical effects of
short sampling time, although further research may be necessary to confirm that.

\section{Conclusions}
Spray formation in a gas-liquid mixing layer is investigated by DNS in the present study. 
To examine whether the simulation fully resolves all the physical scales the mesh resolution 
is varied and the finest mesh consists of about 4 billion cells. The simulation results 
clearly show the detailed processes of how the bulk liquid jet breaks into sheets, then ligaments, 
and finally droplets. The development of the interfacial 
wave is crucial to the sheet formation. Both quasi-2D and fully 3D waves are observed. 
For the 3D waves, the development of the 3D structure is clearly much faster the Rayleigh-Plateau 
instability in the end rim. Ligaments are shown to be generated by fingering 
at end rims of liquid sheets and also by expansion of holes in liquid sheets. The evolution of holes
agrees well with the Taylor-Culick theory. Due to the interaction with the surrounding turbulent 
gas stream, ligaments generally exhibit irregular shapes 
and complex dynamics when they break into droplets. 
The size distributions of droplets in a sampling box at the downstream of the breaking wave 
is investigated for different meshes. It is found that 
a coarse mesh will not only miss the small droplets but also the larger ones. The reason 
is that if the development of the wave is not well resolved
(the mesh is not sufficiently fine for the curvature of the wave crest or the shear layer
above the wave), the sheet formation will be erroneous, resulting
in fake breakups and many tiny ligaments and drops, instead of larger droplets and thicker
ligament as observed in the fine mesh results.  At the end, the Log-Normal and Gamma 
distributions are employed to fit the PDF data and the Log-Normal model seems to 
fit better with the simulation results of the fine meshes. 

\begin{acknowledgements}
This project has been supported by the ANR MODEMI project (ANR-11-MONU-0011) 
program, and the FIRST project supported by 
the European Commission under the 7th Framework Programme 
under Grant Agreement No.\ 265848. This work was granted access to the HPC 
resources of TGCC-CURIE and CINES-Occigen under the allocations 
2015-x20152b7325 made by GENCI. We would also acknowledge PRACE (2014112610)
for awarding us access to CINECA-FERMI and to LRZ-SuperMUC based in 
Italy and Germany. 
\end{acknowledgements}




%

\end{document}